\documentclass[12pt]{article}
\usepackage{epsfig}

\parskip=\medskipamount % space between paragraphs (TeX)
plus.3em minus.5em \abovedisplayshortskip=.5em plus.2em minus.4em
\renewcommand{\thesection}{\arabic{section}}

\catcode`@=11

\@addtoreset{equation}{section} \@addtoreset{equation}{subsection}
\def\theequation{\ifnum\value{section}=0 \arabic{equation}\ignorespaces
\else \ifnum\value{section}=-1 A.\arabic{equation}\ignorespaces
\else \ifnum\value{subsection}=0
\thesection.\arabic{equation}\ignorespaces \else
\thesection.\arabic{subsection}.\arabic{equation}\ignorespaces
                             \fi
                        \fi
                   \fi}

{\catcode`\'=\active \def'{{}^\bgroup\prim@s}}

\catcode`@=12

\newcommand{\bq}{\begin{equation}}
\newcommand{\be}{\begin{equation}}
\newcommand{\fq}{\end{equation}}
\newcommand{\ee}{\end{equation}}
\newcommand{\bqr}{\begin{eqnarray}}
\newcommand{\beqs}{\begin{eqnarray}}
\newcommand{\fqr}{\end{eqnarray}}
\newcommand{\eeqs}{\end{eqnarray}}

\newcommand{\rf}[1]{(\ref{#1})}

\def\bop#1{\setbox0=\hbox{$#1M$}\mkern1.5mu
    \vbox{\hrule height0pt depth.04\ht0
    \hbox{\vrule width.04\ht0 height.9\ht0 \kern.9\ht0
    \vrule width.04\ht0}\hrule height.04\ht0}\mkern1.5mu}
\def\Box{{\mathpalette\bop{}}}                        % box

\begin{document}
\thispagestyle{empty}

\vskip .6in
\begin{center}

{\bf Modification to Special Relativity}

\vskip .6in

{\bf Gordon Chalmers}
\\[5mm]
% {\em address \\
%      address \\
% Los Angeles, CA } \\

{e-mail: gordon@quartz.shango.com}

\vskip .5in minus .2in

{\bf Abstract}

\end{center}

The speed of light is usually taken as one of the fundamental constants.  
String, and field, theories appear to require the altercation of this 
constant into a functional form $E(m,c)$ which is not $E=mc^2$.  The analysis 
requires the re-interpretation of the renormalization group flow 
equations.  There are quantum corrections to the mass-energy relation 
generically for particles of any sort.   A breakdown of special 
relativity follows.  Cosmological data might be one of the best testbeds 
to analyze the computable functional forms of the mass energy relations.

\vfill\break

\noindent{\it Introduction}

Special relativity \cite{SR} has been one of cornerstones of modern physics 
for over a hundred years.  The notion of on-shellness for a particle or string 
is used to define quantum scattering processes through a simple classical 
propagator; propagators dont have to be simple to be consistent with 
basic symmetries however.  The renormalization group has been fully integrated 
with special relativity.  This work is primarily dedicated to theories that 
do not require taking a renormalization limit, such as string models or 
quantum field models that are effective theories.  

A physical interpretation of the renormalization group, together with its 
consistency, 
does require that the energy relation $E(m,c)$ be a function derived in 
a specific quantum field theory; furthermore in the case of light or gravity 
this function tells us that $c(E)$, or that the speed of light varies from 
our currently accepted value $c_0$, the value specified in true vacuum.

The use of the relevant $n$-point functions define the masses and couplings 
through their bare values, 

\bqr 
m^2(m_0^2,\lambda^{(j)}_0; E,c;\Lambda)   \qquad 
\lambda^{(j)}(\lambda^{(j)}_0;m_0^2;E,c,\Lambda) \ , 
\label{rg}
\fqr 
and $\Lambda$ here is the string coupling constant, or a UV cutoff in field 
theory; there could be moduli associated with the compactified space and the 
fluxes which have not been included.  The masses and couplings are fixed 
in energy; if they run, then the running can be specified in the left 
hand side of \rf{rg}.  The latter case is not examined, but could 
be with certain running to match experimental data.  The functional 
form is inverted to find $c(E,m;m,\lambda^{(j)})$.  Of course, higher 
dimension operators may be added to change the functional form of the 
mass-energy relation to model the latter case.  After finding the functional 
form, it is substituted or reshuffled back into the n-point functions 
with higher dimension operators (add and subtract) to determine the full 
scattering.  

These equations usually are called 
renormalization equations \cite{KWJK}, 
and are not used in string theory except for calculations which are 
renormalization special; such as 
three point functions in which phase space requires all momenta to be zero, 
or to cancel IR divergences in the S-matrix rather than then the  
calculation of the full cross section which will cancel IR divergences.  
The equation for the couplings, e.g. 3-pt, 4-pt, $\ldots $, are not 
specifically 
the renormalization group form, as a sum has been performed.  For example, 
the terms 
$A^2\Box A^2$ and $A^2\Box^2 A^2$ have been grouped together in one 
function in $k$-space rather than separated as is typical in RG.  Also, 
note that on-shell requires $k^2=m^2$, by definition.  

The naive substitution of $E=m c_0^2$ might not be consistent with \rf{rg}.  
Note that $c_0$ is used here and in \rf{rg} the correct value $c$ is used.  
An example function is not difficult to find; the bare values of $m_0$ are 
used but the argument can also be obtained with the observed vales $m$.   
Holding $m$, $m_0$ fixed (with $\lambda_0^{(j)}$) could generate a function 
$1=\sin^2 E/c^2/m_0 f(k^2/m_0^2)$ which would generate 
$E=m_0 c^2 g(k^2/m_0^2)$.  This function, as $m_0$ is typically large, varies 
with the momenta.  Although the couplings,
including the masses, are included there is self-consistency that requires the 
speed of light to change.  This is canceled by changing the renormalization 
by scaling the RG point for example.  This scaling of the parameters $m_0$ 
and $\lambda_0^{(j)}$ would restore consistency with special relativity.  
String theory doesn't have a UV renormalization point and the conclusion is 
that the speed of light must be a function of energy.  

The renormalization equations \rf{rg} for the photon and graviton generate the 
equations (with a massless $m=0$), 

\bqr 
c_p(E) \qquad c_g(E) \ .  
\label{speeds}
\fqr 
The h-bar is not used but emphasizes the functional dependence of the speed 
of light and energy, which is quantum due to the flow equations, 

\bqr 
c= \sum c_j E'^j
\label{correction}
\fqr 
in terms of the plank energy $E'=E/\beta$.  The $\beta$ parameter as the 
length scale is not used in terms of compactified dimensions as these 
moduli might flow also.    

The equations in \rf{rg} should be consistent with eachother, including the 
mass and couplings.  Note that these functions $m$ and $\lambda^{(j)}$ 
represent the infinite number of terms of the particular scattering of the 
the vertices with independent couplings.  In the two-point function there 
is one momenta.  The 
entire sum contributes to the relations in \rf{rg}.  

One consequence of the variability of the speed of light is that there are an 
infinite number of quantum corrections to the energy-mass relation, as in 
\rf{correction}.  Another consequence is that special relativity has been 
violated although not in a strong manner because only one dimension, time, 
is essentially treated differently.  This 
implies a special reference frame.  
Furthermore, there may be zeros in $c_p(E)$ and $c_g(E)$; in the case of 
gravity this would require quantization in the pseudo-topological teleparallel 
formalism. 

Each species of particle would have its own $E(c,m)$ relation with 
the zeroth order being its own $E=m_0 c_0^2$ relation.  The similarities 
of the various particles 
would in turn express similarities of their quantization, and the structure 
of spacetime, including the compactification. 

Energy is defined via the field and string theoretic notion of a 
Hamiltonian and its action of flows in time on states.  Varying the 
definition of energy could alter 
the energy mass relation; this could lead to a closer pairing at least 
in practical definition of mass and energy.

The variation of the speed of light or gravity range from zero to very large 
values including infinity, and perhaps is somewhat periodic in energy due 
to a fixed mass parameter and its integer multiples.  
Cosmological 
data might want to be re-examined for clues to these dispersion relations, 
both in light and gravity,  
especially in experimental evidence where there are either large energies, 
such as 
in supernovae, or when energy scales radically fluctuates in short time 
periods 
so that the dispersion can be seen next to a peak along a very steep slope.  
Also, data generated by groups experimentally testing special relativity 
could hold clues within their precision and energy range.

\vfill\break

\end{document}